\documentstyle[preprint,aps,epsf,floats]{revtex}
\begin{document}
\tighten

\def\bfl{{\bbox \ell}}
\def\sss{{\scriptscriptstyle}}
\def\pall{{\parallel}}
\def\bull{\vrule height .9ex width .8ex depth -.1ex}
\def\Dslash{ {D\hskip-0.6em /} }
\def\MeV{{\rm MeV}}
\def\GeV{{\rm GeV}}
\def\Tr{{\rm Tr\,}}
\def\D{{\Delta}}
\def\Ds{{\Delta_6}}
\def\a{{\alpha}}
\def\b{{\beta}}
\def\c{{\gamma}}
\def\d{{\delta}}
\def\m{{\mu}}
\def\M{{\cal M}}
\def\C{{\cal C}}
\def\slash{{\!\not\!}}
\def\nrcpt{NR\raise.4ex\hbox{$\chi$}PT\ }
\def\ket#1{\vert#1\rangle}
\def\bra#1{\langle#1\vert}
\def\ltap{\ \raise.3ex\hbox{$<$\kern-.75em\lower1ex\hbox{$\sim$}}\ }
\def\gtap{\ \raise.3ex\hbox{$>$\kern-.75em\lower1ex\hbox{$\sim$}}\ }
\newcommand{\gsim}{\raisebox{-0.7ex}{$\stackrel{\textstyle >}{\sim}$ }}
\newcommand{\lsim}{\raisebox{-0.7ex}{$\stackrel{\textstyle <}{\sim}$ }}

\def\Journal#1#2#3#4{{#1} {\bf #2}, #3 (#4)}

\def\NCA{\em Nuovo Cimento}
\def\NIM{\em Nucl. Instrum. Methods}
\def\NIMA{{\em Nucl. Instrum. Methods} A}
\def\NPB{{\em Nucl. Phys.} B}
\def\NPA{{\em Nucl. Phys.} A}
\def\PLB{{\em Phys. Lett.}  B}
\def\PRL{\em Phys. Rev. Lett.}
\def\PRD{{\em Phys. Rev.} D}
\def\PRC{{\em Phys. Rev.} C}
\def\PRA{{\em Phys. Rev.} A}
\def\PR{{\em Phys. Rev.} }
\def\ZPC{{\em Z. Phys.} C}
\def\PREP{{\em Phys. Rep.}  }
\def\ANN{{\em Ann. Phys.} }
\def\SCI{{\em Science} }
\def\CJP{{\em Can. J. Phys.}}


\preprint{\vbox{
\hbox{DOE/ER/40561-95-INT00}
\hbox{NT@UW-00-14}
}}
\bigskip
\bigskip

\title{How to Renormalize the Gap Equation\\
 in High Density QCD}
\author{\bf Silas R. Beane$^a$ and  
Paulo F. Bedaque$^b$}

\vspace{1cm}

\address{$^a$Department of Physics and $^b$Institute for Nuclear Theory
\\ University of Washington, Seattle, WA 98195-1560}

\vspace{1cm}

\address{\tt sbeane, bedaque@phys.washington.edu}

\maketitle

\begin{abstract}

We discuss two technical issues related to the gap equation in
high-density QCD: {\it i)} how to obtain the asymptotic solution with well
controlled approximations, and {\it ii)} the renormalization of four-quark
operators in the high-density effective field theory.

\end{abstract}

\vfill\eject

Recently Son obtained the leading exponential behavior of the
superconducting gap in QCD at asymptotically high density using both
an indirect renormalization group argument and a direct QCD
calculation\cite{SON1}. Subsequently many papers confirmed Son's
result\cite{others,pisris,bbs}. However, to the best of our knowledge,
the analytical determinations of the gap suffer from two flaws. First,
the gap equation is divergent and so must be regularized and
renormalized.  Most treatments have taken the baryon density as a
sharp cutoff. This might cause some concern since it leads to the
possibility of contaminating the low energy physics of the gap with ad
hoc high energy physics. Issues of cutoff sensitivity have been
addressed in Ref.~\cite{pisris} and Ref.~\cite{bbs}. Second, the
solution of the gap equation at momenta large compared to the gap has
been obtained by assuming (what appears to be) a particularly
unhealthy approximation which allows the integral equation for the gap
to be expressed as a simple differential equation. The goals of this
paper are modest. Using cutoff regularization we define a renormalized
gap equation and we find the exact asymptotic solution, thus excising
the flaws contained in previous determinations. Our results confirm
Son's original analysis. We also obtain the asymptotic solution using
dimensional regularization with minimal subtraction.

\begin{figure}[t]
\centerline{{\epsfxsize=5.0in \epsfbox{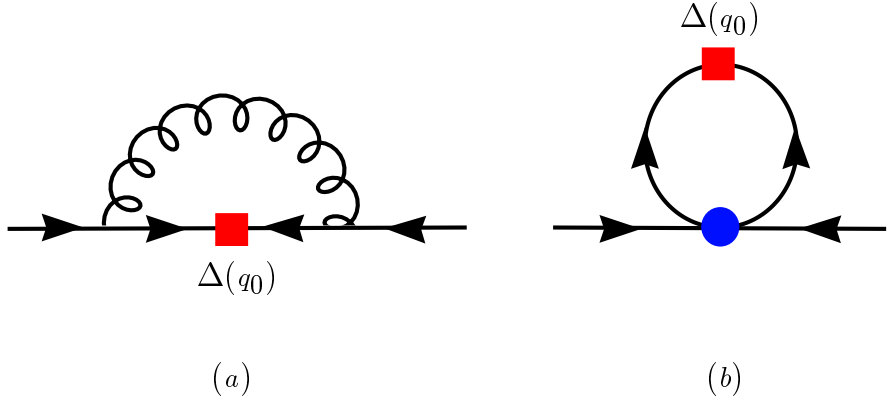}} }
\noindent
\caption{\it
The leading diagrams contributing to the gap equation.  The solid
square denotes the gap, $\Delta$, while the solid circle denotes an
insertion of the counterterm, $\tilde D$.  }
\label{fig:gap}
\vskip .2in
\end{figure}

Many degrees of freedom, including antiparticles and hard gluons, are
not dynamical on the Fermi surface. Hence it is sensible to work with
an effective theory of QCD appropriate to the scales in question.
Explicit construction of the effective field theory appropriate for
momentum scales below $2\mu$, can be found in papers by
Hong\cite{hong}.  The fermions in this effective theory live on the
two-dimensional Fermi surface and so depend only on the parallel
momentum, $q_\parallel$. The gluons on the other hand propagate in
directions perpendicular to the Fermi surface as well and therefore
also depend on the perpedicular momentum, $q_\perp$.  Hence we should
treat the effective field theory as a superposition of two-dimensional
theories, one for each direction on the Fermi surface, interacting
through four-dimensional gluons and contact operators.  Only the
graphs shown in Fig.~(\ref{fig:gap}a) and~(\ref{fig:gap}b) need be
calculated if we are interested in the leading exponential behavior of
the gap.  The sum of these graphs is (after rotating to Euclidean
space)

\begin{eqnarray}
\D (p_\pall) & = & 
\int {{d^2{q_\parallel}}\over (2\pi)^2} {\D(q_\pall )\over{q_\pall^2+\D(q_\pall )^2}}
\Bigg\lbrace {{2g^2}\over 3}\int {{d^2{q_\perp}}\over (2\pi)^2}
\left( {1\over{{\vec q}_\perp^{\;2} +
{{\pi\over 4}M_{\sss d}^2|p_{\sss 0} - q_{\sss 0}|/ |{{\vec q}_\perp}\;|}}}
+{1\over{{\vec q}_\perp^{\;2} + {M_{\sss d}^2}}}\right) + {\tilde D} \Bigg\rbrace,
\label{eq:fullgapcut}
\end{eqnarray}
where the propagators within the parentheses represent
magnetic and electric gluon exchanges, respectively, and

\begin{eqnarray}
M_{\sss d}^2={{{N_f}g^2 \mu^2}\over{2\pi^2}}.
\end{eqnarray}
The effects of hard gluon exchange, antiparticle exchange and any residual
gauge dependence are represented by the coefficient, $\tilde D$, of a
four-fermi operator in the two-dimensional effective theory. 

It is clear that the integration over $q_{\perp}$ is divergent. We
will first regulate the gap equation with a sharp cutoff,
$\Lambda_{\perp}$. In principle there is a cutoff associated with the
parallel integration as well and therefore in general we have ${\tilde
D}={\tilde D}(\Lambda_{\perp},\Lambda_{\pall})$. It is straightforward
to do the integration over $q_{\perp}$ and ${\vec q}_\pall$. We obtain

%
%

\begin{eqnarray}
\D (p_{\sss 0}) & = & 
{1\over 2}{{\cal C}^2}
{\int_{-\Lambda_{\pall}}^{\Lambda_{\pall}}}  {d{q_{\sss 0}}} 
{\D(q_{\sss 0})\over\sqrt{q_{\sss 0}^2+\D(q_{\sss 0} )^2}}
\left(\log {{\cal M}_{\sss \Lambda_{\perp}}\over |p_{\sss 0} - q_{\sss 0}|} 
+{\sss 1\over 2}D(\Lambda_{\perp},\Lambda_{\pall})\right)
\end{eqnarray}
where

\begin{eqnarray}
{\cal M}_{\sss \Lambda_{\perp}}={{4(\Lambda_{\perp})^6}\over{\pi M_{\sss d}^5}};\qquad
{\tilde D}={{g^2D}\over{18\pi}};\qquad{\cal C}={g\over{3\sqrt{2}\pi}}.
\end{eqnarray}
Since $\D (p_{\sss 0})$ is independent of the cutoff, $D$ runs according to

\begin{eqnarray}
{D}(\Lambda_{\perp},\Lambda_{\pall})={D}(\eta,\Lambda_{\pall})
-6\log{{{\Lambda_{\perp}^2}\over{\eta^2}}}.
\end{eqnarray}
Naive dimensional analysis suggests that
${D}(\Lambda_{\perp}={2\mu},\Lambda_{\pall})$ is small and can
therefore be dropped at leading order. The effect of the running of
$D$ appears at next order ~\cite{bbs}.  We then have

\begin{eqnarray}
{\cal M}_{\sss 2\mu}={{2^{10}\sqrt{2}\pi^4\mu}\over{{N_f}^{5/2}g^5}};\qquad
{D}(2\mu,\Lambda_{\pall})=
{D}(\Lambda_{\pall}).
\end{eqnarray}
Dropping all subscripts for simplicity the gap equation becomes

\begin{eqnarray}
\D (p) & = & {1\over 2}{{\cal C}^2}{\int_{0}^{\Lambda}} {d{q}} 
{\D(q)\over\sqrt{q^2+\D(q)^2}}
\left(\log {{\cal M}^2\over |p^2 - q^2|} +D(\Lambda )\right).
\end{eqnarray}
Because of the log singularity at $q\sim p$, the integral is dominated
by momenta $q\sim p$. Therefore in the asymptotic region defined by
$p\gg\D (p)$, we can take $q\gg\D (q)$ under the integral. Hence asymptotically
the gap equation becomes the homogeneous integral equation

\begin{eqnarray}
\D (p) & = & {1\over 2}{{\cal C}^2}{\int_{\Delta}^{\Lambda}} {{d{q}}\over q}
\D(q)\left(\log {{\cal M}^2\over |p^2 - q^2|} +D(\Lambda )\right).
\label{eq:simple}
\end{eqnarray}
Note that although the integral equation is homogeneous, we have by
necessity introduced an infrared cutoff, $\Delta$, which we will see
is related to the overall normalization of the gap.  In order to
proceed\footnote{ The usual way of proceeding is to make the
approximation $\log|p^2-q^2|=
\log{p^2}\theta(p^2-q^2)+\log{q^2}\theta(q^2-p^2)$.  We do NOT make
this approximation in this paper.}, consider the derivative of the gap
equation:

\begin{eqnarray}
\D' (p) & = & {{\cal C}^2}p{\int_{\Delta}^{\Lambda}} {{d{q}}\over q}
{{\D(q)}\over {(q^2 - p^2)}}.
\label{eq:deriv}
\end{eqnarray}
Note that since the counterterm is no longer present, this integral
is convergent as $\Lambda\rightarrow\infty$.

We make an ansatz of the form $\D (p)=p^z$ with $z$ a complex
number. Inserting this solution in eq.~(\ref{eq:deriv}) leads to the
indicial equation:

\begin{eqnarray}
z & = & 
{{\cal C}^2}p^{\sss 2-z}{\int_{\Delta}^{\Lambda}} {{d{q}}}
{q^{\sss z-1}\over {(q^2 - p^2)}}.
\end{eqnarray}
The integral is straightforward to evaluate. It is given by

\begin{eqnarray}
{\it Re}{1\over{ 1-\exp{2\pi i(z-1)}}}
{\oint} {{d{q}}}{q^{\sss z-1}\over {(q^2 - p^2+i\epsilon)}},
\end{eqnarray}
where the contour in the complex q-plane is taken to enclose the poles
at $p-i\epsilon$ and $-p+i\epsilon$ while avoiding the branch point at
the origin. For $\D \ll p \ll\Lambda$ we can take the limits $\D
\rightarrow 0$ and $\Lambda\rightarrow\infty$ since the corrections
are suppressed by powers of $p/\Lambda$ and $\D/ p$.  The resulting
indicial equation is transcendental

\begin{eqnarray}
z & = & {-{\pi\over 2}}{{\cal C}^2}\cot{{\pi z}\over 2}.
\label{eq:indic}
\end{eqnarray}
Therefore the exact asymptotic solution of the gap equation is $p^z$
with z satisfying eq.~(\ref{eq:indic}).  For small z this has the
solution

\begin{eqnarray}
z=\pm i{\cal C}.
\end{eqnarray}
The general asymptotic solution to the gap equation in
the weak coupling limit can then be written as

\begin{eqnarray}
\D (p)=A\sin{({\cal C}\log{\Lambda^*\over p})},
\label{eq:genr}
\end{eqnarray}
where the constants $A$ and $\Lambda^*$ are to be determined.

We will now determine the constant $A$. In the region $\D \leq p <
2\mu$, $\D (p)$ has one maximum, which we will assume is at
$p={\bar\D}$ with $\D\leq{\bar\D}<\Lambda$. This determines $A={\D
({\bar\D})}$. We can find ${\bar\D}$ by plugging the general solution,
eq.~(\ref{eq:genr}), into eq.~(\ref{eq:deriv}) which leads to the
equation

\begin{eqnarray}
0=\D' (p)|_{p={\bar\D}}
={{\cal C}^2}
{\bar\D}\;{\D ({\bar\D})}
{\int_{\Delta}^{\Lambda}} {{d{q}}\over q} \sin{({\cal C}\log{\Lambda^*\over q})}
{1\over {(q^2 - {\bar\D}^2)}}.
\end{eqnarray}
A straightforward computation then gives

\begin{eqnarray}
\cos{x_{\sss \D}}+\sum_{n=1}^\infty {1\over{({\cal C}^2+4n^2)}}(
{\cal C}^2\left[ \exp{\left({n\over{\cal C}}(\pi-2{x_{\sss \D}})\right)}\cos{x_{\sss \D}} 
+\exp{\left(-{n\over{\cal C}}(\pi-2{x_{\sss\Lambda}})\right)}\cos{x_{\sss\Lambda}}\right] 
\nonumber\\
+2{\cal C}n\left[ \exp{\left({n\over{\cal C}}(\pi-2{x_{\sss \D}})\right)}\sin{x_{\sss \D}}
-\exp{\left(-{n\over{\cal C}}(\pi-2{x_{\sss\Lambda}})\right)}\sin{x_{\sss\Lambda}}\right])=0 
\end{eqnarray}
where

\begin{eqnarray}
{x_{\sss \D}}={\cal C}\log{\Lambda^*\over {\D}};\qquad
{x_{\sss\Lambda}}={\cal C}\log{\Lambda^*\over {\Lambda}}.
\end{eqnarray}
The expression under the sum is exponentially
suppressed in the QCD coupling $g$. Therefore, to leading order 
$x_{\sss \D}=\pi/2$ and we can identify ${\bar\D}$ with ${\D}$. This determines
$A= {\D}(\D )$ and the asymptotic solution is

\begin{eqnarray}
\D (p)={\D}(\D )\sin{({\cal C}\log{\Lambda^*\over p})};\qquad\qquad
\D (\D )=\D=\Lambda^* \exp{\left(-{\pi\over{2{\cal C}}}\right)}.
\label{eq:asym}
\end{eqnarray}
We now determine $\Lambda^*$.  We can rewrite eq.~(\ref{eq:simple}) as

\begin{eqnarray}
\D (p) & = & 
{1\over 2}{{\cal C}^2}{\int_{\Delta}^{\Lambda}} {{d{q}}\over q}
\D(q)\log {{\bar{\cal M}}^2\over |p^2 - q^2|} +
{1\over 2}{{\cal C}^2}
\left(D(\Lambda )-D({\bar{\cal M}} )\right)
{\int_{\Delta}^{\Lambda}} {{d{q}}\over q}\D(q)
\label{eq:simple2}
\end{eqnarray}
where

\begin{eqnarray}
{\bar{\cal M}}^2 = {\cal M}^2\exp{(D({\bar{\cal M}} ))}.
\end{eqnarray}
The condition that $\D (p)$ be independent of the choice
of cutoff leads to the renormalization group equation

\begin{eqnarray}
\Lambda {d\over{d\Lambda}}\left(D(\Lambda )-D({\bar{\cal M}} )\right)=
-{\cal C}\tan({\cal C}\log{{\Lambda^*}\over \Lambda})
\left[\left(D(\Lambda )-D({\bar{\cal M}} )\right)+ 
\log{{ {{\bar{\cal M}}^2}\over{\Lambda^2}}}\right]
\end{eqnarray}
where we have used the asymptotic solution, eq.~(\ref{eq:asym}).
The solution of this equation is

\begin{eqnarray}
D(\Lambda )-D({\bar{\cal M}} )=
{2\over{\cal C}}\left[
\tan({\cal C}\log{{\Lambda^*}\over \Lambda})-
{\sin({\cal C}\log{{{\Lambda^*}\over{\bar{\cal M}}}})\over
\cos({\cal C}\log{{\Lambda^*}\over \Lambda})}\right]
-\log{{ {{\bar{\cal M}}^2}\over{\Lambda^2}}}.
\end{eqnarray}
Notice that the difference between counterterms evaluated at different
choices of scale is of order $g^2$.  The counterterm has a strong
cutoff dependence except when $\Lambda\sim{{\bar{\cal M}}}$.  The
peculiar running of the counterterm is very similar to the running of
a counterterm which arises at leading order in effective field theory
treatments of the three-body problem in nuclear
physics\cite{threebod}.

Choosing $\Lambda={{\bar{\cal M}}}$, the renormalized
gap equation is 

\begin{eqnarray}
\D (p) & = & {1\over 2}{{\cal C}^2}{\int_{0}^{{{\bar{\cal M}}}}} {d{q}} 
{\D(q)\over\sqrt{q^2+\D(q)^2}}\log {{{\bar{\cal M}}}^2\over |p^2 - q^2|}.
\label{eq:simple3}
\end{eqnarray}
We can now find $\Lambda^*$ by plugging the general solution,
eq.~(\ref{eq:asym}), into the asymptotic form of the renormalized gap
equation, eq.~(\ref{eq:simple3}), evaluated at $p=\Lambda^*$. We then
have

\begin{eqnarray}
0=\D ({\Lambda^*})
={1\over 2}{{\cal C}^2}{\D ({\D})}
{\int_{\Delta}^{\bar{\cal M}}} {{d{q}}\over q} \sin{({\cal C}\log{\Lambda^*\over q})}
\log {{{\bar{\cal M}}}^2\over |{\Lambda^*}^2 - q^2|}.
\end{eqnarray}
A straightforward computation then gives

\begin{eqnarray}
0=\sin{x_{\sss {\bar{\cal M}}}}-2{x_{\sss {\bar{\cal M}}}}\cos{x_{\sss {\bar{\cal M}}}}
+{1\over 2}\sum_{n=1}^\infty {1\over n}{1\over{({\cal C}^2+4n^2)}}(
{\cal C}^3\exp{\left({2n{x_{\sss {\bar{\cal M}}}}\over{\cal C}}\right)}\cos{x_{\sss {\bar{\cal M}}}}
\nonumber\\
-2{\cal C}^2\left[ 
\exp{\left({2n{x_{\sss {\bar{\cal M}}}}\over{\cal C}}\right)}\sin{x_{\sss {\bar{\cal M}}}}
+n\exp{\left({-n\pi\over{\cal C}}\right)}\right])
\end{eqnarray}
where

\begin{eqnarray}
{x_{\sss {\bar{\cal M}}}}={\cal C}\log{\Lambda^*\over {{\bar{\cal M}}}}.
\end{eqnarray}
The expression under the sum is exponentially suppressed in the QCD
coupling $g$.  Therefore, to leading order ${x_{\sss {\bar{\cal
M}}}}=0$ and we can identify $\Lambda^* ={\bar{\cal M}}$.

The final form for the asymptotic solution is thus

\begin{eqnarray}
\D (p)={\D}(\D )\sin{({\cal C}\log{{\bar{\cal M}} \over p})},
\end{eqnarray}
and the gap is

\begin{eqnarray}
{\D}={{\bar{\cal M}}}\exp{\left(-{\pi\over{2{\cal C}}}\right)}=
{{2^{10}\sqrt{2}\pi^4\mu}\over{{N_f}^{5/2}g^5}}\;
\exp{\left( {D({2\mu},{{\cal M}})\over 2}\right)}\;
\exp{\left(-{3\pi^2\over{\sqrt{2}{g}}}\right)}.
\label{eq:gap}
\end{eqnarray}
This is Son's result\cite{SON1} aside from the additional contribution
to the prefactor from the counterterm, which is expected to be
a number of order one.

One may wonder whether use of cutoff regularization in dense QCD
violates important symmetries. Consistent implementation of
dimensional regularization would erase these concerns.  We will see
that dimensional regularization with minimal subtraction is a quick
way of obtaining the asymptotic solution directly from the gap
equation.  We can continue the four-dimensional measure to an
$2+n$-dimensional measure

\begin{eqnarray}
\int {{d^2q_\pall}\over (2\pi)^2}\int {{d^nq_\perp}\over (2\pi)^n}.
\end{eqnarray} 
The gap equation relevant
to dimensional regularization with minimal subtraction is

\begin{eqnarray}
\D (p_\pall) & = & 
\int {{d^2{q_\parallel}}\over (2\pi)^2} {\D(q_\pall )\over{q_\pall^2+\D(q_\pall )^2}}
\Bigg\lbrace {{2g^2}\over 3}\int {{d^n{q_\perp}}\over (2\pi)^n}
\left( {1\over{{\vec q}_\perp^{\;2} +
{{\pi\over 4}M_{\sss d}^2|p_{\sss 0} - q_{\sss 0}|/ |{{\vec q}_\perp}\;|}}}
+{1\over{{\vec q}_\perp^{\;2} + {M^2}}}\right) + {\tilde D}^{{\sss\overline{MS}}}\Bigg\rbrace.
\label{eq:fullgapdr}
\end{eqnarray}
The integrals of the gluon propagators in $n$-dimensions are

\begin{eqnarray}
\int{{d^n{q_\perp}}\over (2\pi)^n}\;
{1\over{{\vec q}_\perp^{\;2} + {A/ |{\vec q}\;|}}}\;=\;
{1\over{6\pi}}\left({{\lambda^3}\over A}\right)^\epsilon 
\;{{\Gamma \left(\epsilon \right) \Gamma \left(1-\epsilon \right)}\over
{\Gamma \left(1-{{3\epsilon}\over 2} \right)}};
\nonumber\\
\int{{d^n{q_\perp}}\over (2\pi)^n}\;
{1\over{{\vec q}_\perp^{\;2} + {M^2}}}\;=\;
{1\over{6\pi}}\left({{\lambda^3}\over {M^3}}\right)^\epsilon 
\;\Gamma \left({{3\epsilon}\over 2}\right),
\label{eq:drforms}
\end{eqnarray}
where $\epsilon =(2-n)/3$ and $\lambda$ is a renormalization scale.
Absorbing the $1/\epsilon$ pole into the counterterm, we can then
define the regularized gap equation

\begin{eqnarray}
\D (p) & = & {1\over 2}{{\cal C}^2}{\int_{0}^{\infty}} {d{q}} 
{\D(q)\over\sqrt{q^2+\D(q)^2}}
\left(\log {{\cal M}_\lambda^2\over |p^2 - q^2|} +D^{\sss\overline{MS}}(\lambda )\right)
\end{eqnarray}
where

\begin{eqnarray}
{\cal M}_{\sss\lambda}={{4(\lambda )^6}\over{\pi {M_{\sss d}}^5}};\qquad
{\tilde D}^{\sss\overline{MS}}={{g^2D^{\sss\overline{MS}}}\over{18\pi}}.
\end{eqnarray}
This equation was obtained in Ref.~\cite{bbs}.
The counterterm runs according to

\begin{eqnarray}
{D^{\sss\overline{MS}}}(\lambda )={D^{\sss\overline{MS}}}(\eta ) 
-6\log{{{{\lambda}^2}\over{{\eta}^2}}}.
\end{eqnarray}
Physical quantities are $\lambda$-independent so we choose $\lambda =2\mu$. We again consider the
asymptotic gap equation

\begin{eqnarray}
\D (p) & = & {1\over 2}{{\cal C}^2}{\int_{\D}^{\infty}} {d{q}} 
{\D(q)\over q}
\left(\log {{\cal M}_{2\mu}^2\over |p^2 - q^2|} +D^{\sss\overline{MS}}(2\mu )\right).
\end{eqnarray}
Say the asymptotic solution is of the form $\D (p)=p^z$. All scales
then appear multiplied by power law divergences which vanish in
minimal subtraction.  Hence it is appropriate to return to the
unsubtracted expression for the log in eq.~(\ref{eq:drforms}). We then
obtain

\begin{eqnarray}
p^z = {{\cal C}^2}\Gamma (\epsilon )
{\int_{0}^{\infty}} {{d{q}}}{q^{\sss z-1}\over {(q^2 - p^2)^\epsilon}}.
\end{eqnarray}
Again this integral is straightforward to evaluate and leads to

\begin{eqnarray}
p^z = {{\cal C}^2} \Gamma (\epsilon )
\left[ {{p^{z-2\epsilon}}\over 2} \cos{\pi (\epsilon -{z\over 2})}
{{\Gamma \left(\epsilon -{z\over 2}\right) \Gamma \left({z\over 2}\right)}
\over\Gamma \left(\epsilon \right)}\right],
\end{eqnarray}
which as expected reduces to 

\begin{eqnarray}
z = {-{\pi\over 2}}{{\cal C}^2}\cot{{\pi z}\over 2}
\end{eqnarray}
in the limit $\epsilon\rightarrow 0$. Notice that the original $\Gamma
(\epsilon )$ pole from the integration over perpendicular momenta
cancels a $1/\Gamma (\epsilon )$ zero from the integration over $q$.
The asymptotic solution is again

\begin{eqnarray}
\D (p)={\D}(\D )\sin{({\cal C}\log{\Lambda^*\over p})},
\end{eqnarray}
where the prefactor is fixed by the argument presented above.
Now fixing $\Lambda^*$ is trivial since there is only one
scale in the problem. We have

\begin{eqnarray}
\Lambda^*={{\cal M}_{2\mu}}\exp{{D^{\sss\overline{MS}}(2\mu )}\over 2}
\end{eqnarray}
from which follows

\begin{eqnarray}
{\D}={{2^{10}\sqrt{2}\pi^4\mu}\over{{N_f}^{5/2}g^5}}\;
\exp{\left({{D^{\sss\overline{MS}}(2\mu )}\over 2}\right)}\;
\exp{\left(-{3\pi^2\over{\sqrt{2}{g}}}\right)}.
\label{eq:gapdr}
\end{eqnarray}
Evidently the counterterm does not depend on the renormalization scale
in the longitudinal direction in minimal subtraction. As argued
previously, the counterterm is of order $g^{\sss 0}$ and can be
dropped at this order in the expansion.

\vskip0.2in

We thank Martin Savage for valuable conversations.  This work is
supported in part by the U.S. Dept. of Energy under Grants
No. DE-FG03-97ER4014 and DOE-ER-40561.

\vfill\eject

\end{document}